# *Is* Bit *It?*

01001001 01110011 00100000 01000010 01101001 01110100 00100000 01001001 01110100 00111111

*Jennifer L. Nielsen*

## Abstract

In his famous 1989 "it from bit" essay[1], John Wheeler contends that the stuff of the physical universe ("it") arises from information ("bits" – encoded yes or no answers). Wheeler's question and assumptions are re-examined from a post Aspect experiment perspective. Information is examined and discussed in terms of classical information and "quanglement" (nonlocal state sharing). An argument is made that the universe may arise from (or together with) quanglement but not via classical yes/no information coding.

## Introduction

Are we living in a "matrix" ? Is digital information literally all everything comes down to? Or—in other words—is bit "it" ? In a world long post "Gameboy" where the average computer science student is all too aware that simple rules can give birth to complex actualizations, the idea of "it from bit" has a certain "zing" in the zeitgeist –a resonance with the average coffee shop intellectual—that it could not possibly have possessed at the time of Wheeler's origination of the aphorism.

With the recent discovery of a Higgs boson (the particle thought to give our universe mass), we are arguably more up close and personal with whatever "it" is than we ever have been before, and it is more crucial now than ever that we examine closely what we mean by "it" and by "bit" to prepare ourselves for the interpretation of revelations upcoming in an information age in a possible information universe.

## Information

So, what is "information" ? And what is a bit? (It's an easier concept to tackle than the other question—"what is 'it'?"— so let's start here.) While not all of us have encountered Wheeler's 1989 essay, most of us at this point have probably encountered another product of 1989—the Gameboy. Gameboys operate via bits. In a Gameboy game, we may encode a more or less self-consistent, apparently self-contained digital universe such as "pong", the game of "life", Pac man or Pokémon, all using just eight of these things. A bit is a unit of data, generally implemented as an "on" or an "off" signal via a switch which may open or shut in a two-state device, typically represented symbolically in binary notation as 0 and 1. In information theory, we typically define a single bit as the uncertainty of a binary random variable that comes up zero or one with equal probability, or the information / knowledge that is gained when the variable's actual value becomes known. A quantum bit, or "qubit" is a bit more complex as a quantum system that can exist in a superposition of two bit values (it can be on and off at the same time). Everything that pops up on your screen in a Gameboy game, from Mario and Yoshi to a ping-pong ball named Kirby, is written in a program that can be encoded using just 8 bit registers. Furthermore, everything that you perceive around you each day may be represented by neuronal "on/off" switches in your brain and triggered by patterns of on/off signals received from outside events.

    This explanation is satisfying enough at first glance, but is it really helpful? We see right away that the very existence of an "on" or "off", or "yes" and "no" signal implies the existence of a switch which is on or off, or at the very least a question about the physical universe which we can answer with "yes" or "no". We can program a switch into a video game universe—link the open/shut value of a pixelated gate on our game screen to a random number generator—but the switch value itself will ultimately be determined by the state of a physical switch on our machine's silicon motherboard. (The motherboard itself may well fry if we spill water on it—then, bye-bye video game universe. Our universe was self-consistent, yes, at least until dust or water gets on its connectors, but self-contained, perhaps not.) Furthermore in a perfectly functional device the speed that the pixelated Gameboy screen receives the on/off signals from the motherboard is limited physically by the light speed barrier.

Is this really the kind of information that Wheeler is saying that all of us—the "it from bit"—may stem from? We recall that bits don't have to be made out of macroscopic, classical objects; they can also be represented by quantum systems. And we know, in a universe post EPR "spooky action at a distance" and post Alain Aspect's experiment to test for EPR's validity[2], that quantum systems possess another intriguing quality that is sometimes seen by entrepreneuring reality hackers as a potential workaround for the limits of information: nonlocality. Explaining nonlocality, we shall see, via reductionism to something else is much more difficult than is the "bit" concept.

**"Quanglement"**

We know by now via tests of EPR that electrons which have interacted may become entangled. So I'm bored one day in the lab at KU and I mix up a couple spin-entangled electrons and then I bring them home and throw them to my cats, Marko and Zoot, to play with. If Marko's electron is spin up, Zoot's electron is spin down –or vice versa– always, every single time we check, no matter how far away the electrons are from one another. But Marko can't *decide* the spin, and use it to send a Morse or binary code to Zoot if he ships himself off from Lawrence to say Geneva or Las Vegas, because what the spin value *is* is not predetermined by what he does; the answer is given to us randomly by a flip of a coin "outside of time." We can say something about it, encode the spin result into a binary and send the signal digitally to verify that the entanglement is really taking place ("Hey Zoot, your electron really spin down over there? And while we're at it, how are the sardines?"), but that encoded message is limited by the light speed barrier. Here we have encountered two very different forms of "information" – encoded signals, and states of existence correlating *in simul existentia* via entanglement. One is limited by the light barrier, and can be used to convey secondary information about something, post hoc, using bits that we can send via classical signals. The other, while more primary, cannot.

It is obvious now that "information" in the classical bit-encoded sense makes a statement about a thing rather than existing as a thing or "pre-thing" in and of itself; an event occurs and a signal is encoded into an alternate physical medium and sent. Yet in a world where nonlocality is now considered the best explanation for entanglement, it is obvious on some level that the two objects entangled in acausal correlation are involved with one another more profoundly than the two objects exchanging causal info in time via a transactional game of information ping pong. Something important is being shared here, even if we can't directly exploit it.

    Roger Penrose refers to this second type of "information" – information that isn't information – as "quanglement." Quanglement, a word play on quantum entanglement, is used to describe whatever "classical signal free" sharing is going on between the entangled particles. As Penrose puts it, "There is no way to send an ordinary signal by means of quanglement alone."[3] Unlike information conveyed via classical signaling, *quanglement* is not a *code* but rather a reality state *in and of itself*, a state which is shared by multiple particles, in some naïve sense at least outside of time.

    While Penrose and many authors (for example John Cramer of the transactional interpretation) refer to entanglement "signals" going backwards and forwards in time feasibly in order to maintain some sense of cause and effect, for instantaneous entanglement across time and space the entire "signaling" concept is vestigial and recent tests of multi-simultaneity have severely strained if not mortally wounded traditional cause/effect interpretations:

> "Not only are [quantum correlations] independent of the distance, but also it seems impossible to cast them in any real time ordering… *In this sense, quantum correlation is a basic (i.e. primary) concept, not a secondary concept reducible to that of causality between events*: Quantum correlations are directly caused by the quantum state in such a way that one event cannot be considered the 'cause' and the other the 'effect'". [4]

**We Can't Say it All**
Wheeler questioned, optimistically, not just whether "it from bit", but whether we would "someday understand time and space and all the other features that distinguish physics—and existence itself—as the similarly self-generated organs of a self-synthesized information system."[1] We've established that classical information coded into bits cannot be conveyed using quanglement, and that quanglement in and of itself indicates an alternate form of "is-ness" that is not represented adequately by classically conveyed states.

Another possible limit on what we can say using bits may be represented in a proof of Godel's incompleteness theorem. In general, Godel's theorem is summarized by the assertion that any formally generated system capable of expressing elementary arithmetic truth statements cannot be both consistent and complete. As S.C. Kleene put it, "In particular, for any consistent, effectively generated formal theory that proves basic arithmetic truths, there is an arithmetical statement that is true, but not provable in the theory" .[5] If we trust Kleene's interpretation, since binary code represents such a formal system via which we can express elementary truth statements, what we can say with binary code is limited by Godel's theorem.

Say a mad programmer gives Pacman a mini Gameboy within his Gameboy. Any coding language that Pacman may choose to program his mini-Gameboy that sufficiently describes arithmetic is now via definition incomplete. What Pacman can say about himself within his gameworld is limited by what he can code into truth statements; but if everything he programs into his mini Gameboy is consistent within itself, he hasn't completely recreated everything true that he could possibly say. If he tries to completely recreate everything true he can say, his system of language will self-contradict; under Godel's Incompleteness Theorem's most generally accepted interpretation, no encoded language is sufficient to explain itself. From within a Gameboy universe, the Gameboy cannot be entirely encoded and explained.

**It Takes Time to Talk**

Information is often described via Shannon entropy, which quantifies the expected value of information contained in a message. The linking of information to physical entropy alone arguably captures it in a language of evolution in time, while we learn via relativity that any observation of time requires reference to particular ordered physical events. In Vlatko Vedral's

book Decoding Reality, he summarizes the concept of Shannon entropy and how it relates to physical entropy:

> "Physics presents a mathematical formulation of entropy by looking at all the possible states that the system can occupy. Each one of these states will occur with a certain probability that can be inferred from experiments or from some other principle. The logarithm of these probabilities is then taken and the total entropy of the system is then a direct function of this and tells us its degree of disorder:
>
> $S = k \log W$.
>
> "Using the concept of entropy physicists recast the Second Law into the principle that the entropy of a closed system [such as the universe] always increases…Amazingly, this entropy derived by physicists has the same form as the information-theoretic entropy derived by Shannon. Shannon derived his entropy to convey the amount of information that any communication channel can carry. *So in the same sense, maybe we can look at the physicists' concept of entropy as quantifying the information content of a closed system.* The Second Law then simply says that the system evolves to the state of maximal information, where no new information can be contained." – Vlatko Vedral, Decoding Reality [italics mine][6]

In reality, Vedral is restating that information about reality evolves in time as the system evolves. Yet it is in my opinion somewhat circular to look at entropy as quantifying the *information* content of a system – there's no reason to quantify a *quantity*—rather, information content encoded in bits quantifies entropy in an alternate framework based on using the arrangement of physical stuff as bits.

We already determined that bits are states representing information about a system, displayed as symbols but encoded to states in a physical device (such as a Gameboy motherboard), that we can then map to another device (such as a Gameboy interface) using some sort of physical signal. It is fascinating to realize that anything you or I perceive via sight or hearing each day may be represented in on/off neuron switches in our brains/minds. Did we leave the car on? A vibration is emitted into the air—a pattern of sounds representing the car state "on"—and the vibration registers in our ear drum, registering the eardrum vibration state "on" (or more accurately in a specific vibration pattern of "on/offs" we associated with the sound of the car motor from which we infer the car state "on.") We now have information on the physical state of the car. This information travels at the limit of the speed of sound—a physical limit. The ultimate limit on how fast we can get this sort of information across is via the speed of light.

When we say there is a light speed limit on communication, we are saying that it takes time to convey classical information—we are saying that such information is dependent on the motion of physical states in spacetime. It takes time say something, and since what we say depends on the arrangement of physical states in which we are encoding our statements, what we say is governed by the second law as much as anything else. It is thus not only possible but absolutely necessary that classical information is intimately related to how processes evolve in time altogether. Bits are now seen not as a self-generative framework but as taglike descriptors of state changes of pieces within that framework.

We return to Wheeler's speculation of a self-generating information system. While we have seen that bits, and to a greater extent any language made via arrangement of classical information, falls short of offering such a self-consistent and self-created generative structure, we will see that there is still another option.

**It from "Inter-It"**

In the words of Lee Smolin,

> "We can agree that the universe is not identical or isomorphic to a mathematical object, and I've argued that there is no copy of the universe, so there is nothing that the universe is 'like'. What, then, is the universe? Although any metaphor will fail us, and every mathematical model will be incomplete, nonetheless we want to know what the world consists *of.* Not *What is it like*? but *What is it?* What is the substance of the world? We think of matter as simple and inert, but we don't know anything about what matter really *is*. We know only how matter interacts." [7]

I like to tell my summer intro physics students (who are going into medicine), "Physicists like to say that all science is either physics or stamp collecting, but the truth is physics is just dumping the stamps out of the box and plotting how they move and interact.[*] There is a caveat here – in a post EPR, post Aspect experiment world, we know something additional about matter: matter does not just inter-act, it "inter-is." And reality is self-connected in some sense, even when you're not talking about it.

We've argued that bits are just special arrangements of "its" to encode prior knowledge of other "its", and "its" therefore exist before "bits". But we have also seen that quanglement is somehow primary and directly related to this general "itness" in a way that classical information is not.

Is it possible that the universe is not built on a binary but rather is built out of "quanglements" ? We can see via tests of multisimultaneity that quanglement doesn't suffer being younger sister to any event containing her. Is it possible that quanglement and matter always come related hand-in-hand in this way?

Could we live in a universe literally built on entanglement? (Could spacetime or even mass come hand in hand with entanglement itself, or even as a result of it?)

In his award-winning article, "Building Up Spacetime with Quantum Entanglement," Mark Van Raamsdonk argues that the emergence of classically connected spacetimes is correlated with entanglement degrees of freedom in quantum gravity. He argues that classical connectivity in spacetime geometry arises by entangling the degrees of freedom associated with two regions of spacetime, and that disentangling the degrees of freedom results in these regions "pulling apart and pinching off from each other." [8]

Vlatko Vedral has also argued via analogy with the Meissner Effect in superconductors that if the Higgs mechanism for mass generation is proven correct that the resulting Higgs bosons will be found to be entangled, and thus that entanglement and mass itself must be formed concurrently. In an involved discussion of off-diagonal long range order, Vedral maintains that the coherence necessary for mass condensation via the Higgs boson requires entanglement, and that massive objects in the universe thus serve as the necessary witnesses of quantum correlations in the Higgs field.[9] In this scenario, "it" arrives hand-in-hand with "inter-it".

---

[*] Or as I put it to another class, "Physics is not a whole lot like a chick flick – but like chick flicks, physics is all about relationships.

**Conclusion (Is Bit It?)**

So—we're back where we started. Is bit it? Bit in some sense may very well represent what we can detect and manipulate about "it", but quanglement implies something more, a connection that doesn't rely on codified information at all. It's frustrating that we can't get at the "Something More", and poke at it, manipulate it in time, or hack it from the inside—but at least we can be satisfied that bits are representing some*thing.* It is not at least in the classical sense a "put up job"; if we ever find it resembling one, it's generally because we put up the "job" ourselves, and it's always wise to check for what our language is really representing.

> "A hundred years from now, people will look back on us and laugh. They'll say, 'You know what people used to believe? They believed in photons and electrons. Can you imagine anything so silly?' They'll have a good laugh, because by then there will be newer and better fantasies. And meanwhile, you feel the way the boat moves? That's the sea. That's real. You smell the salt in the air? You feel the sunlight on your skin? That's all real. You see all of us together? That's real. Life is wonderful. It's a gift to be alive, to see the sun and breathe the air. And there isn't really anything else."[10] –Michael Crichton, *The Lost World*

> Or, a bit less romantically,

> "a bird is a bird
> slavery slavery
> a knife a knife
> death is death." [11]
> - Zbigniew Herbert

But it's still pretty cool that we can say that in binary.

# References

1. Wheeler, John Archibald. Information, Physics, Quantum: The Search for Links. Proc. 3rd Int Symp. Foundations of Quantum Mechanics, Tokyo (1989)
2. Aspect, Alain, et al. Phys. Rev. Lett. 49, 1804–1807 (1982)
3. Penrose, Roger. The Road to Reality. Alfred A. Knopf NYC (2005)
4. Stefanov, Andre, et al. Quantum Correlations with Spacelike Separated Beam Splitters in Motion: Experimental Test of Multisimultaneity. Physical Review Letters 88, 120404 (2002)
5. Kleene, Stephen Cole. Mathematical Logic. John Wiley & Sons, Inc. NYC (1967)
6. Vedral, Vlatko. Decoding Reality. Oxford University Press. (2010)
7. Smolin, Lee. Time Reborn. Houghton Mifflin Harcourt. NYC (2013)
8. Van Raamsdonk, Mark. Building Up Spacetime with Quantum Entanglement. Gen.Rel.Grav.42:2323-2329 (2010)
9. Vedral, Vlatko. The Meissner Effect and Massive Particles as Witnesses of Quantum Entanglement. arXiv:quant-ph/0410021v1 (2004)
10. Michael Crichton. The Lost World. Knopf. (1995)
11. Herbert, Zbigniew. The Collected Poems: 1956-1998. Harper Collins. (2007)